\documentclass[a4paper,fleqn,usenatbib]{mnras}

\usepackage{footnote}
\usepackage[T1]{fontenc}
\usepackage{ae,aecompl}
\usepackage[flushleft]{threeparttable}
\usepackage{longtable,threeparttablex,booktabs,threeparttable}
\usepackage{longtable}
\usepackage[flushleft]{threeparttable}
\usepackage{longtable,threeparttablex,booktabs,threeparttable}
\usepackage{graphicx}
\usepackage{wrapfig}
\usepackage{pdflscape}
\usepackage{rotating}
\usepackage{epstopdf}
\usepackage{supertabular}
\usepackage{float}
\usepackage[nokeyprefix]{refstyle}
\usepackage{varioref}
\usepackage{xr-hyper}

\usepackage{soul}

\usepackage{graphicx}	
\usepackage{amsmath}	
\usepackage{amssymb}	
\usepackage{multirow,multicol}
\usepackage[T1]{fontenc}
\newcommand{\HI}{H{\sc i}}

\newcommand{\hi}{H{\sc i}~21-cm }
\usepackage{titlesec}
\titlespacing{\section}{1pt}{1.0ex}{1ex}
\titlespacing{\subsection}{1pt}{1.0ex}{1ex}

\title[\hi absorption at $z \approx 3.5$]{uGMRT detection of associated HI 21 cm absorption at $z \approx 3.5$}

\author[J.N.H.S. Aditya et al.]{J. N. H. S. Aditya$^{1}$\thanks{adi.jnhs@gmail.com}, Regina Jorgenson$^2$, Vishal Joshi$^1$, Veeresh Singh$^1$,
\newauthor
Tao An$^3$, Yogesh Chandola$^4$\\
$^{1}$ Physical Research Laboratory, Ahmedabad 380058, India\\
$^{2}$ Maria Mitchell Observatory, Nantucket, Massachussets, USA \\
$^{3}$ Shanghai Astronomical Observatory, Key Laboratory of Radio Astronomy, CAS, 80 Nandan Road, Shanghai 200030, China \\
$^{4}$ Purple Mountain Observatory, Chinese Academy of Sciences, 10, Yuan Hua  Road, Qixia District, Nanjing, 210023, China \\
}

\date{Accepted XXX. Received YYY; in original form ZZZ}

\pubyear{2018}

\begin{document}
\label{firstpage}
\pagerange{\pageref{firstpage}--\pageref{lastpage}}
\maketitle

\begin{abstract}
We report a uGMRT detection of \hi absorption associated with the radio source 8C 0604+728, at $z=3.52965$. The source is at the highest redshift at which associated \hi absorption has been discovered to date, surpassing earlier known absorber at $z \approx 3.39$. We estimate ultraviolet luminosities of $\rm (3.2 \pm 0.1) \times 10^{23}~W~Hz^{-1}$ and $\rm (6.2 \pm 0.2)\times 10^{23}~W~Hz^{-1}$, and ionising photon rates of $\rm (1.8 \pm 0.1) \times 10^{56}~s^{-1}$ and $\rm (5.0 \pm 0.1) \times 10^{56}~s^{-1}$, using data at different epochs; the source shows year-scale variability in both its luminosity and photon rate. The luminosity and photon rate at later epochs are $\approx$6.2 and $\approx $1.7 times higher than thresholds suggested in the literature above which all the neutral hydrogen in the AGN host galaxy is expected to be ionised. The detection demonstrates that neutral hydrogen can survive in the host galaxies of AGNs with high ultraviolet luminosities. We estimate a high equivalent width ratio of 15.2 for the Ly$\alpha$ and HeII emission lines detected in the optical spectrum, that is consistent with AGN photoionisation models. However, a significant contribution from young stellar populations to the excess Ly$\alpha$ flux cannot be ruled out.

\end{abstract}

\begin{keywords}
galaxies  active - quasars  absorption lines - galaxies  high redshift - radio
lines  galaxies
\end{keywords}


\section{INTRODUCTION}

High precision measurements of cosmic neutral hydrogen (\HI) gas density obtained through surveys for damped lyman-$\alpha$ 
(Ly$\alpha$) absorption systems show that the high redshift Universe ($z > 2$) has higher \HI~abundance compared to the local Universe \citep[e.g.][]{bird2017}. Cosmic star formation rate density estimated using infrared and ultraviolet (UV) measurements is also known to peak around redshifts of $z \sim 1.5-3.0$ and decline towards lower redshifts \citep[e.g.][]{gruppioni2013} indicating abundant cold gas at these redshifts. However, searches for \hi absorption associated with active galactic nuclei (AGNs) at $z > 1$ have mostly resulted in non-detections; there are just ten confirmed detections of associated \hi absorption reported to date, at  $z > 1$ \citep[see][and references therein]{chowdhury2020}, compared to over 150 detections at lower redshifts ($z < 1$; \citealp[e.g.][]{vermeulen2003, maccagni2017, aditya2019} ). 
\citet[][]{aditya2018b} find that in their sample of 92 compact flat-spectrum radio sources the \hi absorption strength is lower in the high-redshift ($z \gtrsim 1$) sub-sample. They also estimate detection rates of $28^{+10}_{-8}\%$ and $7^{+6}_{-4}\%$ at low ($z \lesssim 1$) and high redshifts ($z \gtrsim 1$), respectively. However, such a trend is not seen in a sample of 30 Gigahertz-Peaked Spectrum (GPS) sources studied by \citet[][]{aditya2018a}. Two detections are reported in a sample of six sources at $z > 1$, tentatively yielding a high detection rate of $33^{+43}_{-21}\%$ (the large error is due to the small sample size), which is consistent with the rate at low redshifts. The detection rate in the GPS sources at $z > 1$ appears to be higher than that of the flat-spectrum sources, although more observations would be needed to confirm this.  

In the case of low-redshift ($z < 1$) extended radio sources, \citet[][]{maccagni2017} estimate a detection rate of $(16.0 \pm 6.8) \%$, which is lower than $(32 \pm 7.9) \%$ for the compact sources in their sample. The difference could be due to a low gas covering factor in extended radio sources, that leads to a low measured optical depth \citep[see e.g.][]{pihlstrom2003, curran2013b}. At high redshifts, however, extensive searches for associated \hi absorption towards extended radio sources are not yet available.

In the case of flat-spectrum radio sources, a redshift evolution in the gas properties and selection bias in the AGN samples, where the high-$z$ sources typically have higher UV and 1.4 GHz radio luminosities due to the Malmquist effect, have been proposed as the possible causes for the low detection rate in samples at high redshifts \citep[e.g.][]{curran2012a, aditya2016, aditya2018b}. A high 1216 \AA~UV luminosity would ionize the neutral gas in AGN surroundings \citep[e.g.][]{curran2012a}, while a high 1.4 GHz radiation would raise the spin temperature of the gas \citep[e.g.][]{field1959, bahcall1969}; both the effects lead to a lowering of the \hi absorption strength. Here, luminosity at 1216 \AA~is considered instead of at 912 \AA~where the photons ionize the gas, since mere excitation of the gas above the ground state by the Ly$\alpha$ photons is enough to cease the 21-cm absorption \citep[e.g.][]{curran2008}.  It is currently not clear which of the three causes viz. a redshift evolution, and AGN 1216 \AA~UV and 1.4 GHz radio luminosities have a dominant effect in lowering the associated \hi absorption strength, and thus, reducing the detection rate in high-$z$ high-luminosity systems with compact radio sources \citep[e.g.][]{aditya2018b, curran2019}.

\begin{figure}

\includegraphics[height=9cm, width=9cm]{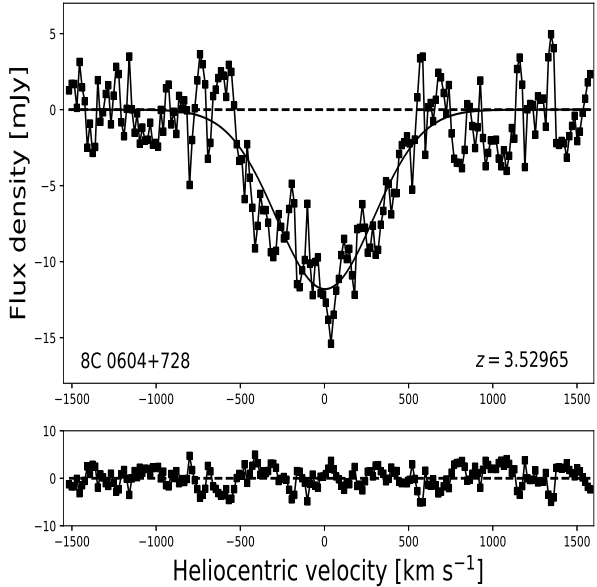} \\
\vspace*{-7mm}
\caption{The H{\sc i}~21-cm absorption spectrum towards 0604+728, at $z = 3.52965$. The spectrum is obtained after combining present uGMRT observations with those reported in \citet[][]{aditya2016}. The top panel shows a Gaussian fit to the absorption feature, while the bottom panel shows the residual after the fit is subtracted from the spectrum.}\label{det}
\end{figure}

\citet[][]{curran2012a} have proposed a cut-off rest-frame UV luminosity of $\rm L_{UV} = 10^{23}~W~Hz^{-1}$ above which all the neutral hydrogen in a Milky Way-like host galaxy is expected to be ionized. They pointed out that no associated \hi absorption has been detected where the AGN luminosity is higher than $\rm 10^{23}~W~Hz^{-1}$. However, \citet[][]{aditya2017} have reported a detection towards TXS 1954+513, at $z = 1.223$, that has $\rm L_{UV}= (4.1 \pm 1.2) \times 10^{23}~W~Hz^{-1}$; this is the first case of \hi detection associated with a quasar with $\rm L_{UV} \gtrsim 10^{23}~W~Hz^{-1}$, albeit with a large error on the luminosity \citep[see also][]{curran2019}. The detection of \hi absorption indicates that hydrogen could indeed `survive' in the host galaxies of AGNs with high UV luminosities, exceeding the above threshold. In this paper we report the upgraded Giant Metrewave Radio Telescope (uGMRT) detection of \hi absorption associated with the flat-spectrum radio source 8C 0604+728, at $z = 3.52965$. Earlier, \citet[][]{aditya2016} reported a tentative detection towards the source. Our re-detection of the feature presented in this paper confirms that it is not caused by transient RFI, and it improves the signal-to-noise ratio of the original detection reported by \citet[][]{aditya2016}, leading to a firm detection of \hi absorption. The source is at the highest redshift at which \hi absorption has been discovered to date, surpassing the previously known intervening and associated absorbers at $z \approx 3.39$, towards PKS 0201+113 and TXS 0902+343, respectively \citep[][]{kanekar2007, uson1991}. This is also the second case of \hi absorption associated with a quasar with $\rm L_{UV} > 10^{23}~W~Hz^{-1}$.


\begin{figure}
\centering
\begin{tabular}{c}
    \includegraphics[width=9.0cm, height=8cm]{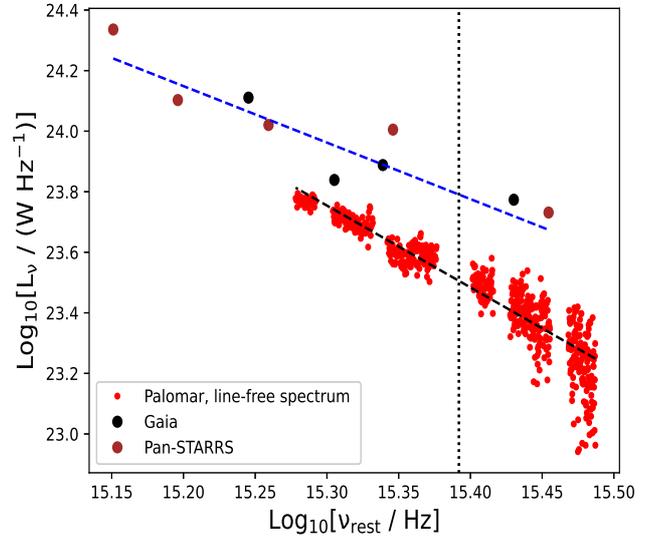} \\
\end{tabular}
\vspace*{-3mm}
\caption{The blue and black dashed lines are power-law fits to the Gaia and Pan-STARRS data, and Palomar line-free spectrum, respectively. The vertical dotted line corresponds to 1216 \AA.}\label{powerlaw}
\end{figure}

\section{OBSERVATIONS}

\subsection{Radio observations}
The uGMRT observations towards the target were conducted on 1st December 2018, using the 250--500 MHz receiver, and with the GMRT software correlator as the backend. We used a bandwidth of 4.17 MHz centred at 313.6 MHz, and subdivided into 512 channels, which provided a velocity resolution of $\rm \approx 7.8 \ km \, s^{-1}$ and a coverage of $\rm \approx 3994 \ km \ s^{-1}$. The standard calibrator, 3C~147, was observed to calibrate the flux density scale and the system bandpass, while 0410+769 was observed for phase calibration. The on-source time was $\sim 198$ mins. 

The data were analysed in the Astronomical Image Processing System (AIPS), using standard procedures described in \citet[][]{aditya2016}. The source has a flux density of $(1790.2 \pm 0.1)$ mJy and is unresolved in the final continuum image. We note that although the measurement errors are small, the typical error on the flux density scale, dominated by systematics and calibration, is expected to be $\lesssim 10\%$. An absorption feature is detected in the Stokes I spectrum (see Figure~\ref{append}), that has similar peak flux density, $\approx -17$ mJy, in RR and LL stokes. In order to improve the signal-to-noise ratio, the spectrum is combined with the earlier GMRT spectrum reported in \citet[][]{aditya2016}, using inverse-variance weighting, after normalising the flux density (see Figure~\ref{det}). The integrated optical depth of the absorption feature is $\rm 4.69 \pm 0.15 \ km \ s^{-1}$, which is consistent with $\rm 4.29 \pm 0.28 \ km \ s^{-1}$ reported in \citet[][]{aditya2016}, within $1\sigma$ error.

\subsection{Optical spectrum and redshift}  
\citet[][]{jorgenson2006} reported the optical observations towards the source using the 200-inch Hale telescope of the Palomar observatory, conducted in the years 1996 and 1997. Observations were made with a $1\arcsec$ slit, with the double spectrograph, resulting in a $4-6$ \AA~ spectral resolution at 4000 \AA, with a coverage of $\sim 4000 - 9000$ \AA. Further details on the observations are provided in \citet[][]{jorgenson2006}. The data were reduced using standard IRAF packages \citep[see][]{jorgenson2006}.

We used the ultraviolet composite AGN spectrum by \citet[][]{shull2012} to identify the metal lines $\lambda~1240.1$ \AA~NV, $\lambda~1303.5$ \AA~[O I], $\lambda~1404.8$ \AA~[O IV], $\lambda~1486.5$ \AA~[N IV], $\lambda~1640.5$ \AA~[HeII], $\lambda~1660.8, 1666.2$ \AA~[O III] and $\lambda~1750.2$ \AA~[N III] in the spectrum (see Figure~\ref{append}). By fitting Gaussian profiles to the emission lines we estimate the redshift of the source to be $3.52965 \pm 0.00051$ which is consistent with the redshift $z = 3.53$ reported by \citet[][]{jorgenson2006}. Here, the Gaussian fitting errors are propagated to estimate the error.

\section{DISCUSSION}
\subsection{\hi absorption}
The 1.4 GHz VLA continuum image reported by \citet[][]{taylor1996}, with a spatial resolution of $\approx 10$ kpc, shows strong emission from two connected radio lobes (with peak flux densities of $\approx 831$ mJy and $\approx 79$ mJy), where the weaker lobe possibly represents a jet component extending towards North-West from the core \citep[see also][for description]{aditya2016}. The projected size between the radio lobes is $\approx 35$ kpc \citep[][]{taylor1996}. The milliarcsec-scale Very Large Baseline Interferometry (VLBI) image at 5 GHz shows the resolved structure of the stronger lobe that consists of multiple compact and diffuse emission components, with flux densities in the range of few$\times 10$ mJy to $\approx 121$ mJy \citep[][]{britzen2007}. The separation between the outermost components in the VLBI image is $\approx 157$ pc. Overall, the radio structure resembles a core-jet, where the core possibly lies at the eastern end of the VLBI emission structure. 

A single Gaussian profile is fitted to the \hi absorption as shown in Figure~\ref{det}; the reduced $\chi ^{2}$ value is 1.05. The Gaussian has a Full Width at Half-Maximum (FWHM) of $\rm 676 \pm 31~km\ s^{-1}$, an amplitude of $-11.8 \pm 0.1$ mJy, and a peak at $\rm 3.6 \pm 13.1~km~s^{-1}$ that is consistent with the AGN redshift, implying that the absorber is associated with the AGN. We note that the line width is wide compared to the typical line width ($\rm \approx 200~km~s^{-1}$) of associated absorbers in the literature \citep[e.g.][]{curran2016b}. The feature could be arising due to absorption from a single gas cloud, or due to a convolution of several profiles arising from multiple absorbers with large velocities. In the latter case, the absorption could be arising either against a few or all of the radio continuum structural components revealed in the milli-arcsec and arcsec-scale images. The former case would mean that the gas is obscuring only a fraction of the background radio continuum, yielding a smaller gas covering factor ($f < 1$, see \citealp[][]{curran2013b}). Since the radio source is unresolved in the GMRT beam, the above case would mean that the measured \hi optical depth could be a lower limit. VLBI observations at redshifted \hi frequencies would be helpful in locating the absorbers at parsec scales.

\subsection{AGN luminosity and ionising photon rate}\label{lum_ion}

As discussed above, it has been proposed that the \hi absorption strength could be critically dependent on the AGN 1216 \AA~luminosity \citep[e.g.][]{curran2008, curran2012a}. \citet[][]{curran2012a} have proposed a cut-off 1216 \AA~luminosity of $\rm 10^{23}~W \ Hz^{-1}$ above which they point out that \hi absorption associated with AGN has never been detected. Using models with parameters similar to that of the Milky Way the authors estimated that all the neutral gas in the AGN host galaxy would be ionized where the AGN luminosity is higher than this cut-off. Since the source 8C 0604+728 is at a high redshift, $z \approx 3.5$, we expect a high AGN luminosity, owing to the Malmquist selection bias, as the source is a part of the flux-limited Caltech-Jodrell Flat-spectrum sample \citep[][]{taylor1996}. To test the above hypothesis, we, therefore, estimated the AGN 1216 \AA~UV luminosity using the optical spectrum of the source. First, we corrected the optical spectrum for Galactic dust extinction using magnitudes towards the target taken from \citet[][]{schlegel1998}. A power-law fit was used to interpolate to all the wavelengths in the spectrum. Next, we considered the metal line-free part of the spectrum to estimate the UV continuum, as shown in the left panel of Figure~\ref{powerlaw}. A power-law was fitted to the luminosities where higher weight was given to data with better noise characteristics. The luminosity at 1216 \AA~was estimated to be $\rm L_{UV} = \rm (3.2 \pm 0.1) \times 10^{23} \ W \ Hz^{-1}$. 

Quasars are known to show high flux density variability in their UV and optical wavebands, over a timescale ranging from days to years \citep[e.g.][]{punsly2016}. Since the optical spectrum was observed in 1996--1997, in order to test the possible variability at later epochs, we collected the optical data from the recent Gaia \citep[][]{prusti2016} and Pan-STARRS DR2 \citep[][]{flewelling2016} surveys, to estimate the AGN luminosity (see Table~\ref{lit_data}). These observations were conducted during the years 2010 to 2015. After correcting for the galactic extinction, and fitting a power-law to the data (see Figure~\ref{powerlaw}), we estimate the luminosity at 1216 \AA~to be $\rm (6.2 \pm 0.2) \times 10^{23}~W\ Hz^{-1}$. Clearly, both our luminosity estimates are higher than the threshold, $\rm 10^{23}~W\ Hz^{-1}$, albeit there is a year-scale variability.

Further, using gas models with parameters similar to the Milky Way, \citet[][]{curran2012a} also propose a cut-off ionising photon rate of $\sim 3 \times 10^{56}$ photons per second from the AGN, a threshold above which, again, no detection of \hi absorption has been made. This is estimated by computing the left-hand side of the Equation~\ref{em} (see \citealp[][]{curran2013}) for the ultraviolet spectrum of the quasar. The Equation~\ref{em} represents the equilibrium between photoionisation and recombination in a hydrogen gas nebula, where $\rm L_{\nu} $ is the specific luminosity, $\rm h$ is the Planck constant, $\rm n_{p}$ and $\rm n_{e}$ are the proton and electron densities, $\alpha_{\rm A, B}$ is the radiative recombination rate coefficient of hydrogen, $\rm r_{ion}$ is the radius up to which the gas is ionised, and $\rm \nu_{ion}$ corresponds to 1216 \AA, i.e.  $\rm 10^{15.39}~Hz$. 

\begin{equation}\label{em}
\rm \int_{\nu_{ion}}^{\infty} \frac{L_{\nu}}{h \nu} d\nu = 4 \pi \int_{0}^{r_{ion}} n_{p} n_{e}\alpha_{\rm A, B}r^{2}dr 
\end{equation}

The estimation of the ionising photon rate for a quasar involves the integration of the specific luminosity (factored by $\rm h \nu$) over all wavelengths shorter than 1216 \AA. To test if the ionising photon rate in the current source exceeds the above threshold, we estimate the photon rate (Q) using power-law fits to the Palomar optical spectrum, and the Gaia and Pan-STARRS data (following the description in \citealp[][]{curran2013}); the estimates are $\rm (1.80 \pm 0.1) \times 10^{56}\ s^{-1}$ and $\rm (5.0 \pm 0.1) \times 10^{56}\ s^{-1}$, respectively. Although the estimate using the Palomar data is lower, the photon rate inferred using the recent literature data is $\approx 1.7$ higher than the threshold ionising photon rate of $\rm 3 \times 10^{56}~s^{-1}$.

We note that the distribution of the neutral gas around the AGN could be quite complex. While several observations suggest that the gas could be more concentrated near the AGN core, multiple cases are available where the absorption is distinctively detected against the extended radio jets or near the terminal hot-spots \citep[e.g.][]{peck1999, morganti2013, aditya2018a}. While the ionisation models studied by \citet[][]{curran2012a} would apply in the former case, where the gas in the near vicinities of the core is critically affected by the ionising radiation, \hi absorption could still be detected from the gas that is more distant from the nucleus. The current source 8C 0604+728 could be one such case, where the absorption could be occurring against bright radio-emitting components detected in the VLBI/VLA studies that are located away from the nucleus. Alternatively, the absorption could be arising from pockets of dense neutral gas where the gas density is higher than $\rm n = 10~cm^{-3}$ \citep[see e.g.][]{curran2016}, the canonical value used in the models of \citet[][]{curran2012a}.

We note that the ionising photon rates are estimated by extrapolating power-law fits to wavelengths shorter than $\sim 912$ \AA. Further, luminous high-$z$ quasars are known to show high UV variability, even $\gtrsim 200 \%$, on year timescales \citep[e.g.][]{punsly2016}, as in the current case, where the 1216 \AA~luminosity is varying by a factor of $\approx 2$ over a period of $\gtrsim 14$ years. These factors could critically affect our interpretation of the ionising thresholds, and need to be included in the AGN gas-ionisation models.        
             
\begin{figure}
  \begin{tabular}{c}
    \includegraphics[width=9cm, height=4.5cm]{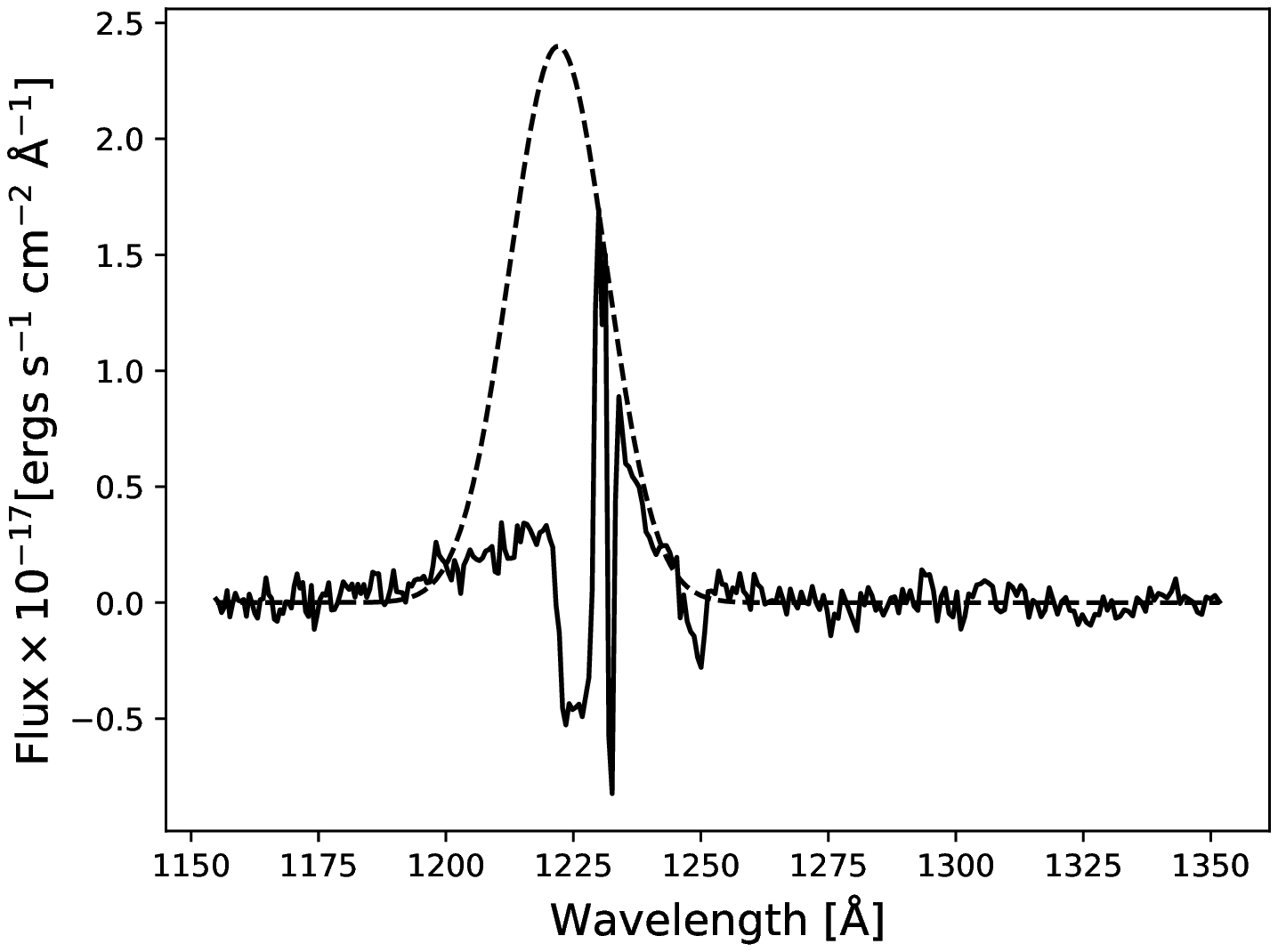} \\
    \includegraphics[width=9cm, height=4.5cm]{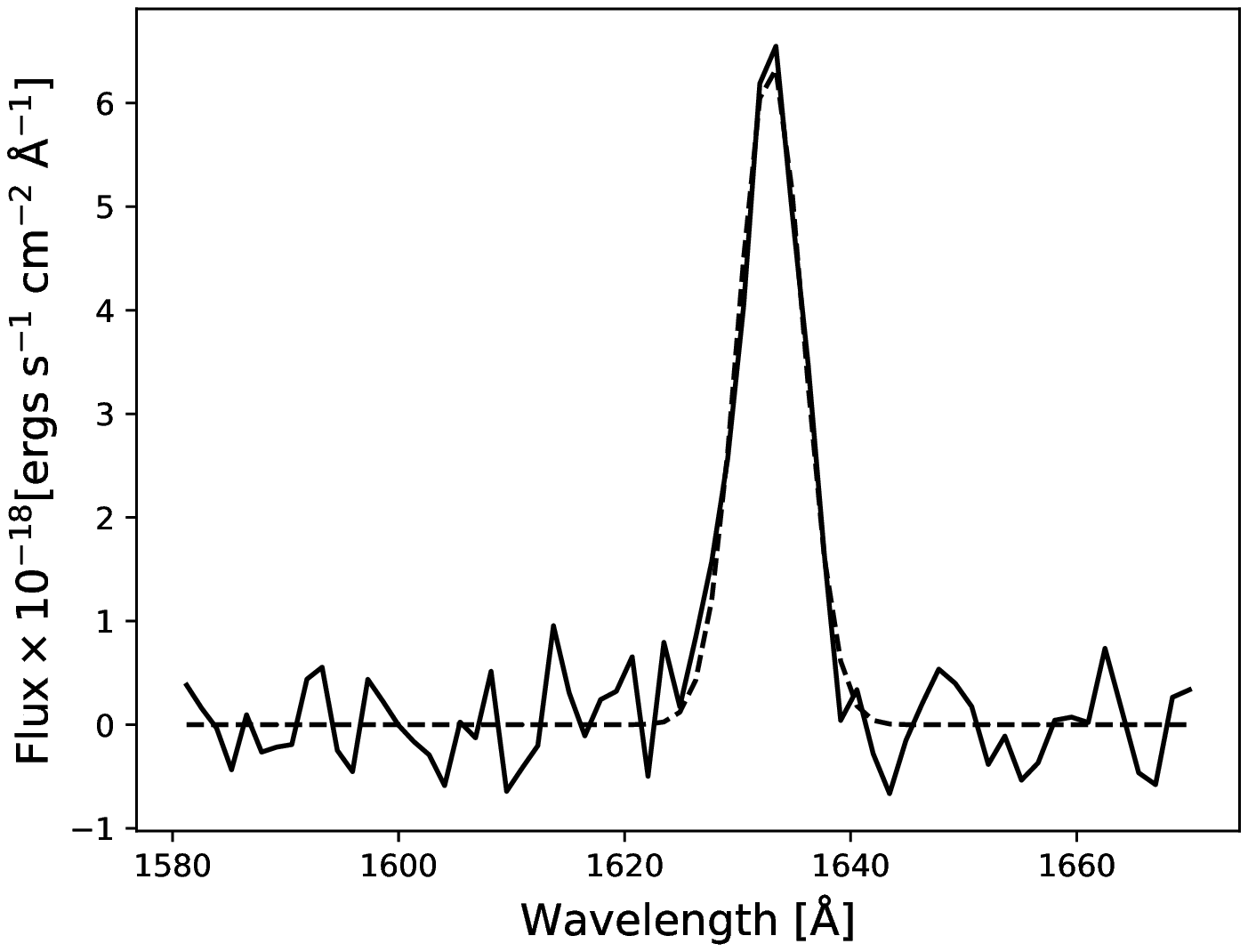} \\

  \end{tabular}
  \caption{Gaussian fits to the continuum subtracted rest-frame Ly$\alpha$ (top panel) and HeII profiles (bottom panel), shown with dashed lines.}\label{gauss_fits}
\end{figure}

\subsection{Ly$\alpha$ and $\rm HeII$ emission}
The optical spectrum shows complex Ly$\alpha$ absorption and emission features (see Figure~\ref{gauss_fits}). Two strong absorption features, a narrow and a broad one, are visible in the spectrum (see Figure~\ref{gauss_fits}), where the broad feature is relatively blueshifted. Assuming that the underlying Ly$\alpha$ emission line has a Gaussian profile (see e.g. \citealp[][]{vanojik1997}), we fit a Gaussian with an FWHM of 22.3 \AA~to the spectrum in the source rest frame. Overall, a high fraction of the Ly$\alpha$ flux, and a significant continuum emission, appear to be attenuated by \HI~absorption. While the narrow and broad absorption features could be caused by strong \HI~absorption, the overall depression in the Ly$\alpha$ flux could be caused by the presence of a significant amount of dust \citep[e.g.][]{vanojik1997}.

A high fraction ($>60\%$) of Ly$\alpha$ emitters at high redshifts, that harbour a compact radio source ($\rm < 50~kpc$), are indeed known to show deep absorption troughs in their Ly$\alpha$ emission \citep[e.g.][]{vanojik1997}. These studies also find a strong correlation between distortion in the Ly$\alpha$ velocity field and the structure of the radio source, as can be seen in the current case, implying an association between the radio jets and the gas. Statistics of quasar absorption lines also show that the chance of a strong \HI~absorber lying in the small redshift interval of the Ly$\alpha$ emission line ($\Delta z \sim 0.03$) is just $\approx 2\%$ \citep[e.g.][]{petitjean1993}, which means that the absorber is most likely associated with the AGN host galaxy. The absorption troughs in the Ly$\alpha$ spectrum, thus, provide another evidence for the presence of dense neutral gas likely associated with an AGN, with ultraviolet luminosity $\rm L_{UV} > 10^{23}~W~Hz^{-1}$.

The ionisation potential of $\rm He^{+}$ (54.4 eV) is higher than that of $\rm H^{0}$ (13.6 eV), thus, the production of a notable {\rm HeII} 1640 \AA~flux relative to Ly$\alpha$ requires either a source with a hard ionising spectrum (like the AGN) or significant collisional ionisation by shocks \citep[e.g.][]{humphrey2019}. On the other hand, neutral hydrogen can be excited by a number of factors, for example, photoionisation by AGN, soft ionising photons from stellar populations, collisional excitation, shocks, etc. \citep[see e.g.][]{villarmartin2007}. The ratio of the Ly$\alpha$ and {\rm HeII} lines is thus a powerful diagnostic of the nature of ionising sources in the galaxy and the physics of the gaseous nebulae. \citet[][]{villarmartin2007} find that in their sample of high redshift radio galaxies, the $z > 3.0$ sources systematically show higher Ly$\alpha$ / HeII equivalent width (EW) ratios compared to the galaxies at $2 < z < 3$, with a typical EW ratio $> 15$ in $z > 3$ systems. They argue that the excess Ly$\alpha$ flux is produced due to excess star formation in the $z > 3$ systems. However, \citet[][]{humphrey2019} compute AGN photoionisation models and find that the radiation field of luminous AGNs can as well explain the excess Ly$\alpha$ / HeII ratios. Further, these models could be used to deduce constraints on the physical parameters, like the ionisation parameter (U), gas chemical abundance ($Z$), etc. For the current source, we derive the Ly$\alpha$ / HeII EW ratio, by shifting the spectra to the rest-frame and fitting Gaussians to the emission profiles. The EW of Ly$\alpha$ and HeII are 76.1 \AA~and 5.0 \AA~respectively, yielding a ratio of 15.2, consistent with those of the $z > 3$ systems in \citet[][]{villarmartin2007} sample. The parameters in the models of \citet[][]{humphrey2019} that are closest to the EW ratio correspond to a high ionisation parameter, $\rm -2 < log_{10}U < -1$, and high metallicity, $Z / Z_{\odot} = 1.0$. The presence of strong {\rm HeII} emission and the consistency with the AGN photoionisation models means that the AGN is likely the dominant source of ionisation. However, due to the high chemical abundance, a significant contribution to the Ly$\alpha$ emission from mechanisms such as stellar photoionisation cannot be ruled out. 

Using the relation $\rm N_{H^{+}} / N_{H^{0}} = 10^{5.3} \times U$ \citep[e.g.][]{davidson1979} for the ratio of number density of ionized and neutral hydrogen atoms (ionisation fraction), the constraints on U yield the limits $\rm 10^{3.3} < N_{H^{+}} / N_{H^{0}} < 10^{4.3}$, implying that much of the gas in the Ly$\alpha$ emitting nebulae is ionized. Thus, the \hi absorption could likely be arising from
gas clouds located in regions mentioned in Section~\ref{lum_ion}, and/or from the outer Ly$\alpha$ absorbers detected in the optical spectrum.  

\section{SUMMARY}
We have conducted uGMRT observations to detect \hi absorption associated with a flat-spectrum radio source, 8C 0604+728 at $z = 3.52965$. The absorber is at the highest redshift to date surpassing the previously known intervening and associated absorbers at $z \approx 3.39$, in the literature. We estimate an AGN ultraviolet luminosity and an ionising photon rate of $\rm (6.2 \pm 0.2) \times 10^{23}~W~Hz^{-1}$ and $\rm (5.0 \pm 0.1) \times 10^{56}~s^{-1}$, respectively, which are $\approx 6.2$ and $\approx 1.7$ times higher than the thresholds suggested by \citet[][]{curran2012a} above which all the neutral hydrogen in the host galaxy is expected to be ionized. However, we note that the AGN shows high variability in both 1216 \AA~UV luminosity and ionising photon rate (by factors of $\approx 2$ and $\approx 3$, respectively), over year timescales. The detection demonstrates that neutral hydrogen could survive in the surroundings of AGN with high ultraviolet luminosities. We estimate a high Ly$\alpha$/HeII equivalent width ratio of 15.2, which can be explained by models of photoionisation either by the AGN or by young stellar populations. A strong {\rm HeII} emission in the optical spectrum suggests that the AGN could be the dominant ionising source. The photoionisation models predict a high ionisation parameter, $\rm -2 < log_{10}U < -1$, and a high gas metallicity, $Z / Z_{\odot} = 1.0$ for the Ly$\alpha$ emitting gas nebulae.

\section*{ACKNOWLEDGEMENTS}
We thank the anonymous referee for a very helpful report that improved the quality of the paper.
TA thanks the financial support from the National Key R\&D Programme of China (2018YFA0404603). YC thanks
Center for Astronomical Mega-Science, CAS, for FAST distinguished young researcher fellowship (19-FAST-02). We thank the staff of the GMRT who have made these observations possible. The GMRT is run by the National Centre for Radio Astrophysics of the Tata Institute of Fundamental Research. This research has made use of the NASA/IPAC Extragalactic Database (NED), which is funded by the National Aeronautics and Space Administration and operated by the California Institute of Technology.
\section*{DATA AVAILABILITY} 
The data from GMRT \hi observations are available in the online archive, at https://naps.ncra.tifr.res.in/goa/data/search with proposal codes 15NKa01 and $35\_027$. The data related to optical observations will be provided upon request to Regina Jorgenson (rjorgenson@mariamitchell.org).      
\bibliographystyle{mnras}
\bibliography{ms}
\newpage
\appendix
\section{APPENDIX}

\begin{table}
\footnotesize
\caption{Photometric data from literature. The Pan-STARRS and Gaia observations were conducted during the years 2010-2015.} 
\label{lit_data}
\begin{small}
\begin{tabular}{cccccc}
\hline 
Flux            &  error\_flux  & Survey    & Wavelength & Band  & Ref. \\
$\mu$Jy         & $\mu$Jy &           &      \AA~  &       &            \\

\hline
 
 69.3     &   5.8 &    Pan-STARRS  &  9596.4   &      y           &  1    \\
 39.2     &   2.4 &    Pan-STARRS  &  8652.0   &      z           &  1   \\    
 38.3     &   1.9 &    GAIA        &  7724.6   &    $\rm G_{RP}$  &  2        \\  
 30.7     &   1.0 &    Pan-STARRS  &  7479.8   &      i           &  1   \\
 19.3     &   1.0 &    GAIA        &  6729.9   &    $\rm G_{RP}$  &  2     \\
 20.8     &   0.2 &    GAIA        &  6226.2   &    $\rm G_{BP}$  &  2     \\
 27.0     &   1.7 &    Pan-STARRS  &  6125.7   &      r           &  1     \\ 
 14.1     &   2.1 &    GAIA        &  5046.2   &    $\rm G_{BP}$  &  2      \\
 12.3     &   0.8 &    Pan-STARRS  &  4772.2   &      g           &  1     \\
\hline
\hline
\end{tabular}
\end{small}
\begin{tablenotes}
\item
References: (1) \citet[][]{flewelling2016}, (2) \citet[][]{prusti2016}. 
\end{tablenotes}

\end{table}

\hspace{-8cm}

\begin{figure*}

\begin{tabular}{cc}
    \includegraphics[height=0.4\linewidth, width=0.5\linewidth]{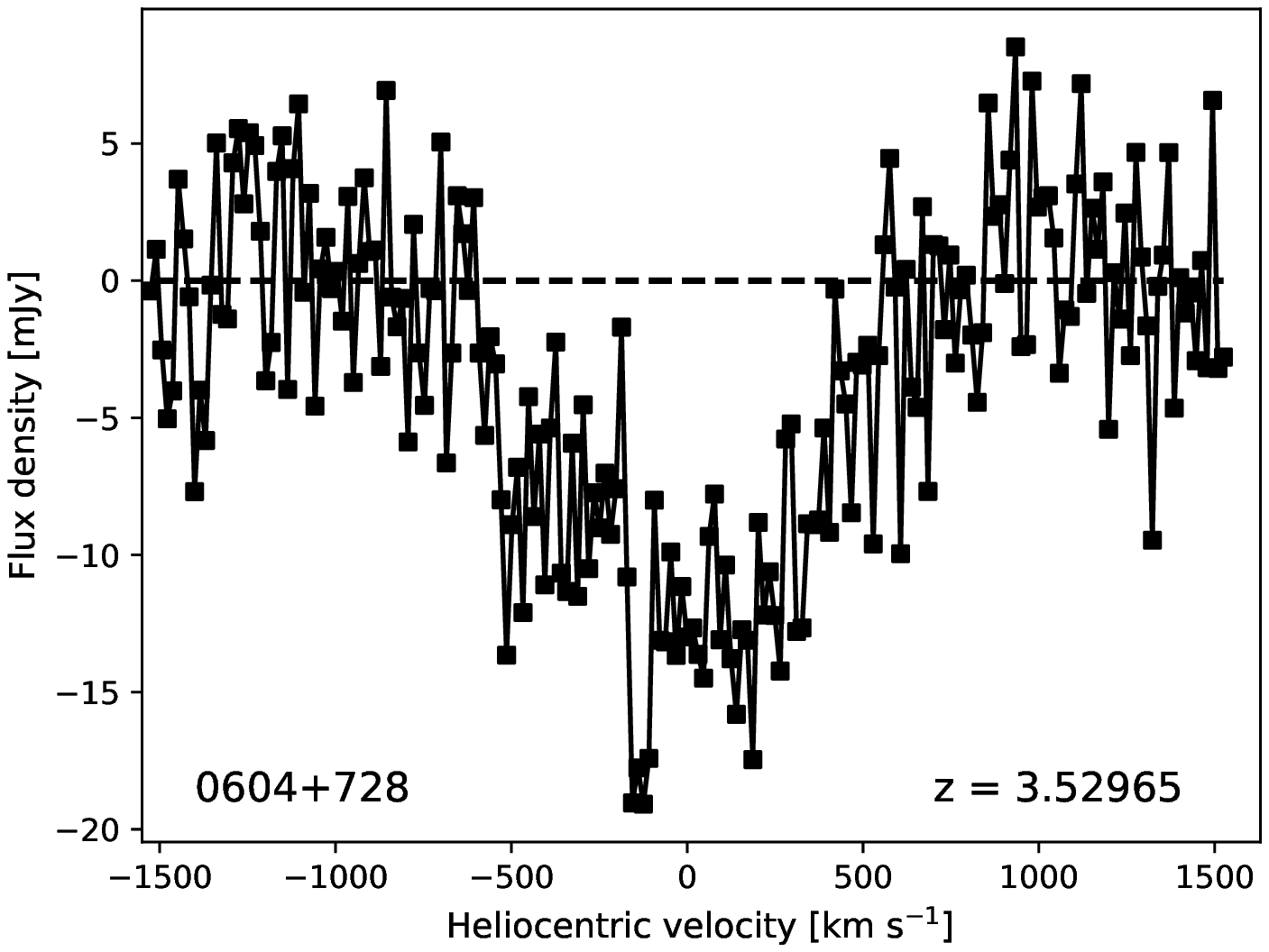} &
    \includegraphics[height=0.4\linewidth, width=0.5\linewidth]{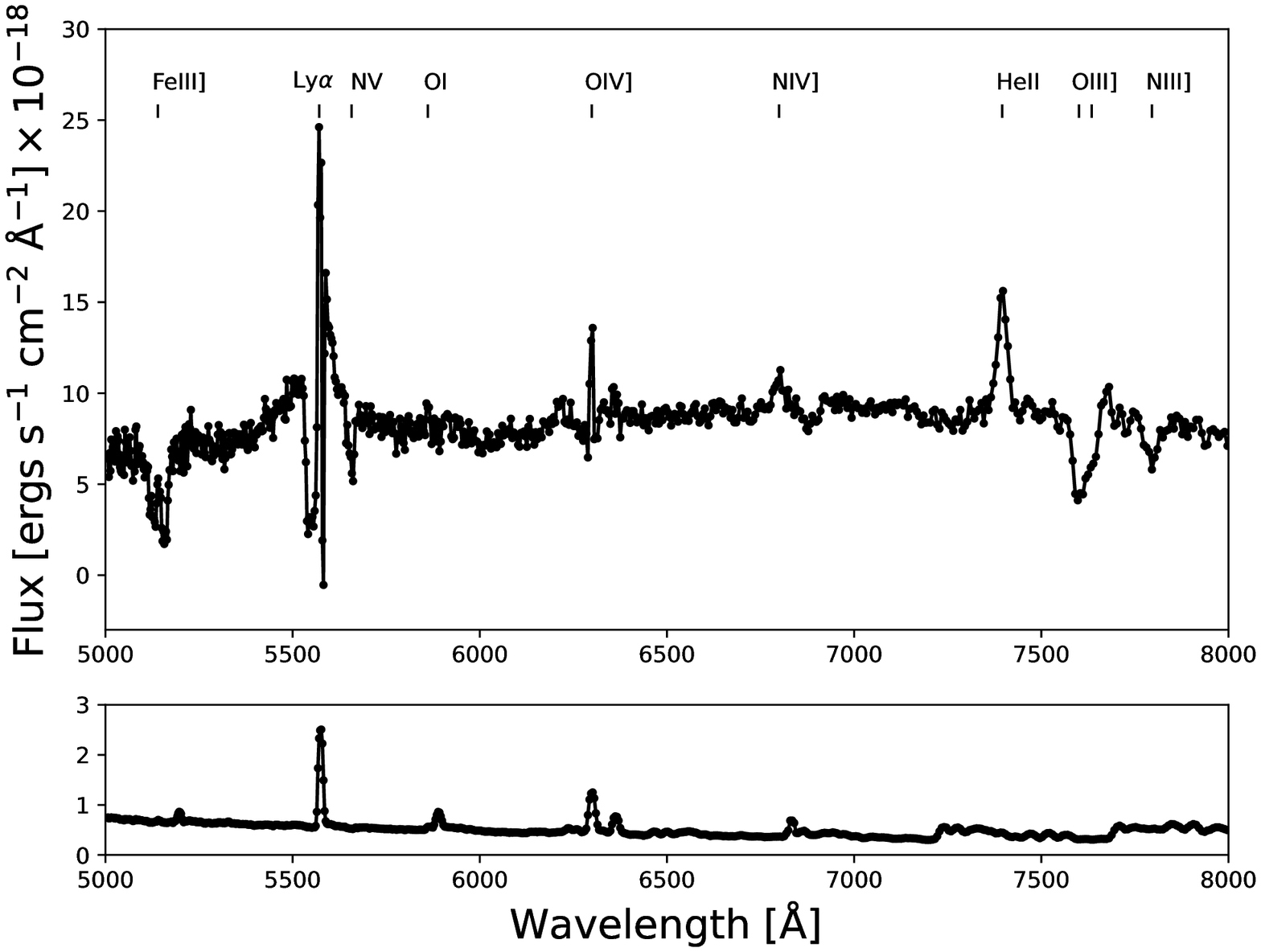} \\
\end{tabular}
\caption{Top panel: uGMRT H{\sc i}~21-cm absorption spectrum from our observations in 2018. Bottom panel: Optical spectrum of the source, with flux errors, obtained from Palomar observations in 1996-1997.}\label{append}
\end{figure*}

\label{lastpage}

\end{document}